\documentclass[10pt,twocolumn,letterpaper]{article}

\usepackage{iccv}
\usepackage{times}
\usepackage{epsfig}
\usepackage{graphicx}
\usepackage{amsmath}
\usepackage{amssymb}
\usepackage{color}
\usepackage{algorithm}  
\usepackage{algorithmic}
\usepackage[table,xcdraw]{xcolor}
\usepackage{caption}
\usepackage{threeparttable}
\usepackage{url}
\usepackage[accsupp]{axessibility}
\usepackage[pagebackref=true,breaklinks=true,letterpaper=true,colorlinks,bookmarks=false]{hyperref}

\iccvfinalcopy % *** Uncomment this line for the final submission

\def\iccvPaperID{6677} % *** Enter the ICCV Paper ID here

\ificcvfinal\fi

\begin{document}

%%%%%%%%% TITLE
\title{ Dense Deep Unfolding Network with 3D-CNN Prior for Snapshot Compressive Imaging}
\author{Zhuoyuan Wu$^\dagger$
\qquad
Jian Zhang$^{\dagger,\ddagger}$
\qquad
Chong Mou$^\dagger$ \\
$^\dagger$Peking University Shenzhen Graduate School, Shenzhen, China\\
$^\ddagger$Peng Cheng Laboratory, Shenzhen, China \\
\tt\small {wuzhuoyuan@pku.edu.cn; zhangjian.sz@pku.edu.cn; eechongm@gmail.com}
}
% \author{Zhuoyuan Wu$^\dagger$\\
% % Peking University\\
% % {\tt\small wuzhuoyuan@pku.edu.cn}
% % For a paper whose authors are all at the same institution,
% % omit the following lines up until the closing ``}''.
% % Additional authors and addresses can be added with ``\and'',
% % just like the second author.
% % To save space, use either the email address or home page, not both
% \and
% Jian Zhang$^{\dagger,\ddagger}$\\
% % Institution2\\
% % First line of institution2 address\\
% % {\tt\small secondauthor@i2.org}
% \and
% Chong Mou$^\dagger$
% \and
% $^\dagger$Peking University Shenzhen Graduate School, Shenzhen, China\\
% $^\ddagger$Peng Cheng Laboratory, Shenzhen, China
% }
\maketitle
\let\thefootnote\relax\footnotetext{This work was supported in part by National Natural Science Foundation of China (61902009). (\textit{Corresponding author: Jian Zhang.}) }
% Remove page # from the first page of camera-ready.
\ificcvfinal\thispagestyle{empty}\fi

%%%%%%%%% ABSTRACT
\begin{abstract}
Snapshot compressive imaging (SCI) aims to record three-dimensional signals via a two-dimensional camera. For the sake of building a fast and accurate SCI recovery algorithm, we incorporate the interpretability of model-based methods and the speed of learning-based ones and present a novel dense deep unfolding network (DUN) with 3D-CNN prior for SCI, where each phase is unrolled from an iteration of Half-Quadratic Splitting (HQS). To better exploit the spatial-temporal correlation among frames and address the problem of information loss between adjacent phases in existing DUNs, we propose to adopt the 3D-CNN prior in our proximal mapping module and develop a novel dense feature map (DFM) strategy, respectively. Besides, in order to promote network robustness, we further propose a dense feature map adaption (DFMA) module to allow inter-phase information to fuse adaptively. All the parameters are learned in an end-to-end fashion. Extensive experiments on simulation data and real data verify the superiority of our method. The source code is available at \href{https://github.com/jianzhangcs/SCI3D}{https://github.com/jianzhangcs/SCI3D}.
% The code is available at \url{https://github.com/masterwu2115/SCI2020
%   The ABSTRACT is to be in fully-justified italicized text, at the top
%   of the left-hand column, below the author and affiliation
%   information. Use the word ``Abstract'' as the title, in 12-point
%   Times, boldface type, centered relative to the column, initially
%   capitalized. The abstract is to be in 10-point, single-spaced type.
%   Leave two blank lines after the Abstract, then begin the main text.
%   Look at previous ICCV abstracts to get a feel for style and length.
\end{abstract}

%%%%%%%%% BODY TEXT
\section{Introduction}

As a major branch of compressive sensing (CS) \cite{duarte2008single, zhao2014image, zhao2018cream}, snapshot compressive imaging (SCI) develops for the aim of capturing high dimensional signals such as video \cite{llull2013coded} or spectral data \cite{wagadarikar2009video}. Incorporating additional hardware components in the imaging system, SCI samples consecutive video frames or spectral channels and integrates sampled data to get compressed measurements. Then original sampled signals will be reconstructed from such measurements through various algorithms. Building encoder and decoder in a hardware plus software fashion, limitation on recording high dimensional signals in terms of bandwidth and memory have been overcome by a large margin. In this paper, we concentrate on video SCI.

Taking measurements comprising information on sampled signals and the sensing matrix as input, SCI recovery algorithms focus on how to solve an ill-posed inverse problem mathematically. Conventional methods employ prior knowledge of image or video as regularizers and solve such a sparsity-regularized optimization problem iteratively \cite{yuan2016generalized, yang2014video, maggioni2012video, reddy2011p2c2}. Though superior in convergence and theoretical analysis, these algorithms tend to have complex computation, and it is quite difficult to select appropriate priors, which restricts the practical application of these methods. Thanks to the vigorous development of deep learning, several SCI reconstruction methods based on convolutional neural networks (CNN) have been proposed in recent years, which prefer to learn a direct mapping from measurements to sampled frames \cite{yuanbirnat, ma2019deep, han2020tensor, meng2020gap, iliadis2018deep, yoshida2018joint, yuan2020plug}. Compared to traditional ones, learning-based methods can achieve more decent results while lacking interpretability.

\begin{figure*}[!t]
\centering 
% \subfigure[]{
\includegraphics[width=0.9\textwidth]{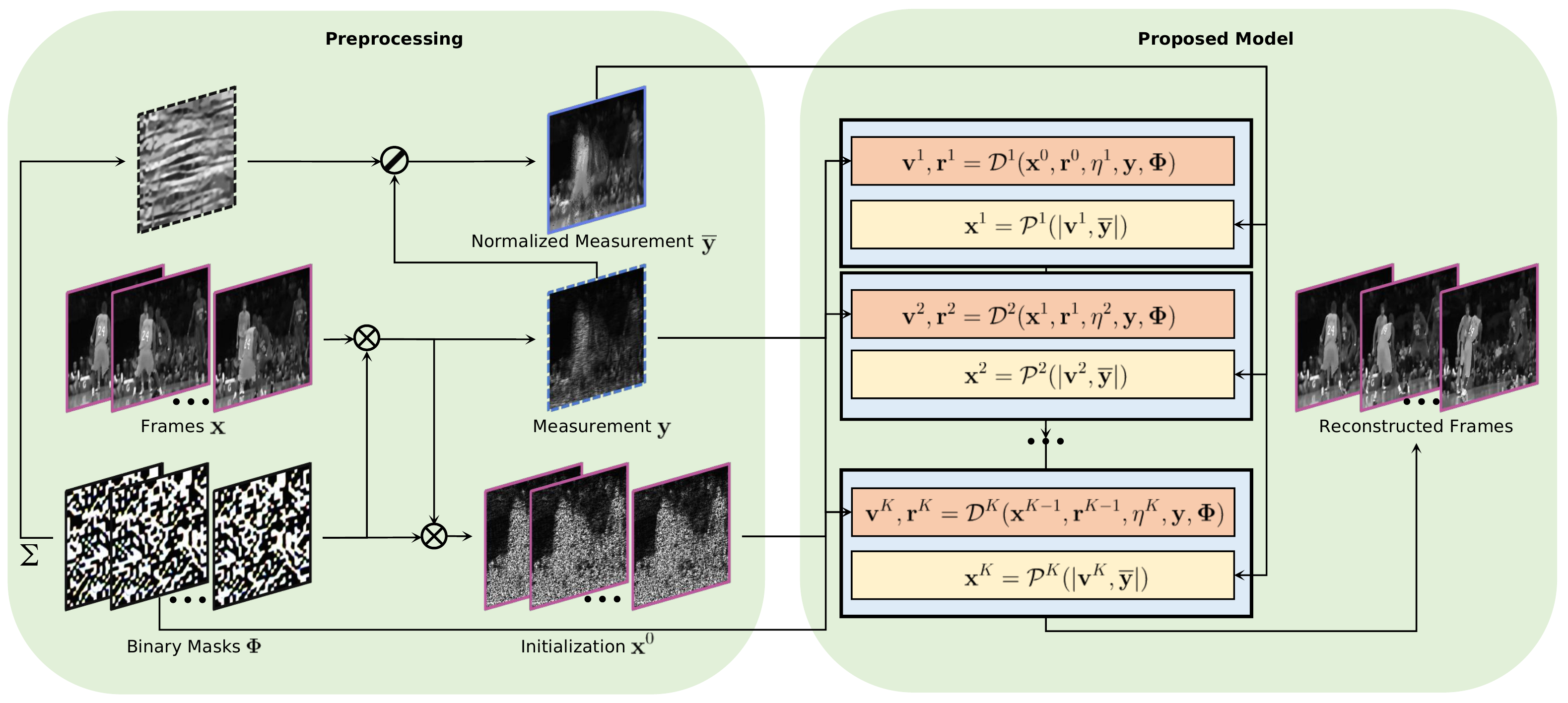}
% \subfigure[]{
% \includegraphics[width=1\textwidth]{detail_cropped_compressed.pdf}
% }
\caption{
Illustration of the video SCI (left) and our proposed method (right). \textbf{Left}: a sequence of video frames $\mathbf{x}$ are passed through dynamic masks $\boldsymbol{\Phi}$ and then compressed to the measurement $\mathbf{y}$. Normalized measurement $\overline{\mathbf{y}}$ is acquired by element-divide the sum of dynamic masks. $\mathbf{x}^0$ is a coarse reconstruction of multiple frames. \textbf{Right}: details of our proposed network comprising $K=10$ phases where each iteration of HQS update step is unrolled to a phase including data module $\mathcal{D}^k$ and prior module $\mathcal{P}^k$.
}
\label{SCI}
\end{figure*}

Combining the merits of both model-based and learning-based methods, we design a dense deep unfolding network with 3D-CNN prior by unrolling Half-Quadratic Splitting \cite{geman1995nonlinear} into a deep network for optimizing the SCI recovery problem. As depicted in Fig.~{\ref{SCI}}, each iteration of HQS is mapped to a phase of our proposed network. Unlike previous learning-based methods that construct 2D-CNN to model images, we design an effective 3D-CNN that is more suitable to exploit spatial-temporal correlation between frames. Besides, to decrease the information loss between neighboring phases in deep unrolling networks (DUN) caused by up-sampling and down-sampling in channel dimension, we propose to fuse corresponding dense feature map of adjacent phases. Moreover, to make information selectively transmitted through phases, we propose a dense feature map adaption module. Our contribution can be summarized as below:

% {\color{red}explain the reason why using 3D CNN rather than 2D one and inspiration of long-range information integration.} 
\begin{itemize}
    \item We build a novel dense deep unfolding network with 3D-CNN prior for SCI recovery. Our method achieves a significant improvement both visually and qualitatively compared with current state-of-the-art methods.
    \item We first incorporate 3D-CNN as priors into the proximal mapping in deep unfolding network, which better characterizes the spatial-temporal correlations among frames.
    \item We propose dense feature map (DFM) fusion to break the limitation on the transmission of network information, which brings another increment of 0.45dB.
    \item We present a dense feature map adaption (DFMA) module to adaptively fuse DFM, which helps to transmit useful information between phases. 
    % The whole branches of DFMA bring another gain of 0.32dB on the basic of DFM fusion.

\end{itemize}

\section{Related Work}
\subsection{Snapshot compressive imaging}
Model-based methods serve structure sparsity of sampled video as a regularizer. For example, total variance \cite{yuan2016generalized}, sparsity \cite{maggioni2012video, zhao2016video}, gaussian mixture model \cite{yang2014video}, optical flow \cite{reddy2011p2c2} and non-local low rank \cite{liu2018rank} is employed as prior knowledge. Although model-based methods can be directly applied to different sensing matrices without retraining, the main drawback of such algorithms is that the performance is highly restricted by the selection of prior. Moreover, iteratively solving optimization problems can be time-consuming.

In recent years, researches based on deep learning have surged in computational imaging \cite{zhang2018ista, zhangopine, you2021}. Some concentrate on constructing neural networks to learn a mapping from compressed measurements to sampled frames. For example, deep fully-connected network \cite{iliadis2018deep} is proposed to learn the linear mapping, a method of jointly optimizing the exposure patterns with the reconstruction framework \cite{yoshida2018joint} considering the constraints enforced by hardware. BIRNAT \cite{yuanbirnat} is designed where a CNN reconstructs the first frame, and the other frames are reconstructed by a bidirectional dense neural network. Most Recently, MetaSCI \cite{wang2021metasci} is composed of a shared backbone that has light-weight meta-modulation parameters for different masks. Others engage in building up deep unrolling networks, which map iterations of optimization algorithms to phases of DUN. Various methods like Tensor-ADMM \cite{ma2019deep}, Tensor-FISTA \cite{han2020tensor}, GAP-Net \cite{meng2020gap}, PnP-FFDNet \cite{yuan2020plug}, Anderson-Accelerated unrolled network \cite{li2020end} and STEP-SCI \cite{wu2021spatial} have been proposed for SCI. Nonetheless, previous DUN-based methods have three potential drawbacks: i) they tend to rely on 2D-CNNs to excavate the spatial-temporal correlation; ii) majority methods convey information that contains video frames from one to another phase, lacking more potential information; iii) when it comes to fusing inter-phase information, there is not an adaptive strategy using only useful information. Bearing the above concerns in mind, in this paper, we put forward a new design of unrolling network and show how to make use of spatial-temporal correlation and decrease information loss effectively.

\subsection{3D Convolutional Neural Networks}
3D-CNN is good at capturing spatial-temporal correlation, especially for tackling video processing tasks. Over the years considerable applications of 3D-CNN are proposed in such areas as action recognition \cite{ji20123d, tran2015learning}, event classification \cite{teivas2017video}, video super-resolution \cite{kim20183dsrnet, luo2020video}, medical image segmentation \cite{cciccek20163d}, video inpainting \cite{wang2019video, chang2019free} and so on. However, only a few pieces of research on SCI recovery problems refer to 3D-CNN. Until lately, consisting of multi-group reversible 3D convolutional neural networks, RevSCI \cite{cheng2021memoryefficient} is presented for large-scale video SCI, where several 3D convolutions are incorporated into their proposed networks of feature extraction, reversible mapping, and reconstruction. Nevertheless, shallow architecture restricts its performance due to limited receptive field. In this paper, we design an elaborate network for video SCI recovery to address such issues, which exhibits a more effective ability to utilize the spatial-temporal correlation.

\section{Mathematical Model of SCI}
Fig.~{\ref{SCI}} gives a brief depiction about the pipeline of video SCI system. Consider there is a measurement $\mathbf{Y}\in \mathbb{R}^{1\times H\times W}$ with the dimension of $H\times W$ which is compressed from a $B$-frame video sequence $\{\mathbf{X}_i\}_{i=1}^B\in \mathbb{R}^{1\times H \times W}$ through $B$ masks $\{\boldsymbol{\Phi}_i\}_{i=1}^B \in \mathbb{R}^{1\times H \times W}$, such procedure can be expressed as bellow:
\begin{equation}
    \mathbf{Y} = \sum_{i=1}^B \boldsymbol{\Phi}_i \otimes \mathbf{X}_i + \mathbf{N}, \label{1}
\end{equation}
where $\mathbf{N} \in \mathbb{R}^{1\times H\times W}$ denotes the noise and $\otimes$ is the Hadamard (element-wise) product. Eq.~{(\ref{1})} is equivalent to the following linear form:
\begin{equation}
    \mathbf{y} = \boldsymbol{\Phi} \mathbf{x} + \mathbf{n}, \label{2}
\end{equation}
where $\mathbf{y} = \text{vec}(\mathbf{Y}) \in \mathbb{R}^{HW}$ is the vectorized measurement, $\mathbf{n} = \text{vec}(\mathbf{N}) \in \mathbb{R}^{HW}$ is the vectorized noise and $\mathbf{x} = \text{vec}(\mathbf{X}) = [\text{vec}(\mathbf{X}_1)^\top, ..., \text{vec}(\mathbf{X}_B)^\top]^\top \in \mathbb{R}^{HWB}$ are the vectorized frames. The sensing matrix $\boldsymbol{\Phi} \in \mathbb{R}^{HW\times HWB }$ can be expressed as a block diagonal form as below:
\begin{equation}
    \boldsymbol{\Phi} = [\text{diag}(\text{vec}(\boldsymbol{\Phi}_1)), ..., \text{diag}(\text{vec}(\boldsymbol{\Phi}_B))]. \label{3}
\end{equation}
The compression rate is equal to $1/B$. The reconstruction error of SCI is bounded even when $B>1$, which is proved in \cite{jalali2019snapshot}. The purpose of SCI recovery problem is to reconstruct video frames $\mathbf{x}$ given the measurement $\mathbf{y}$ and masks $\boldsymbol{\Phi}$, where $\boldsymbol{\Phi}$ is often to be binary. Thus, $\boldsymbol{\Phi} \boldsymbol{\Phi}^\top$ is assumed invertible, commonly in a diagonal form with a straightforward calculation of inversion.

\section{Proposed Method}

\subsection{Measurement Energy Normalization}
\label{Measurement Energy Normalization}
As proposed in \cite{yuanbirnat}, a strategy of measurement energy normalization is designed to introduce more motion information which assists in reconstructing. The normalized measurement $\overline{\mathbf{Y}}$ is defined by
\begin{equation}
    \overline{\mathbf{Y}} = \mathbf{Y} \oslash \boldsymbol{\overline{\Phi}},
\end{equation}
where $\oslash$ is the Hadamard (element-wise) division and $\boldsymbol{\overline{\Phi}} = \sum_{i=1}^B \boldsymbol{\Phi}_i$ is the sum of $B$ sensing matrices which denotes how many pixels in $\{\mathbf{X}_i\}_{i=1}^B$ are integrated into the measurement $\mathbf{Y}$. Equivalently, $\overline{\mathbf{y}} = \text{vec}(\overline{\mathbf{Y}}) \in \mathbb{R}^{HW}$ is the vectorized normalized measurement. Generally speaking, the range of pixel values in $\mathbf{Y}$ can be broad, making it less effective to concatenate a non-energy-normalized measurement into the network directly. As depicted in Fig.~{\ref{SCI}}, visually more pleasant than $\mathbf{Y}$, $\overline{\mathbf{Y}}$ can be served as a coarse reconstruction result which preserves a certain amount of motion information.

\begin{figure*}[!t]
\centering 
% \subfigure[]{
% \includegraphics[width=1\textwidth]{SCI_cropped.pdf}}
% \subfigure[]{
\includegraphics[width=0.9\textwidth]{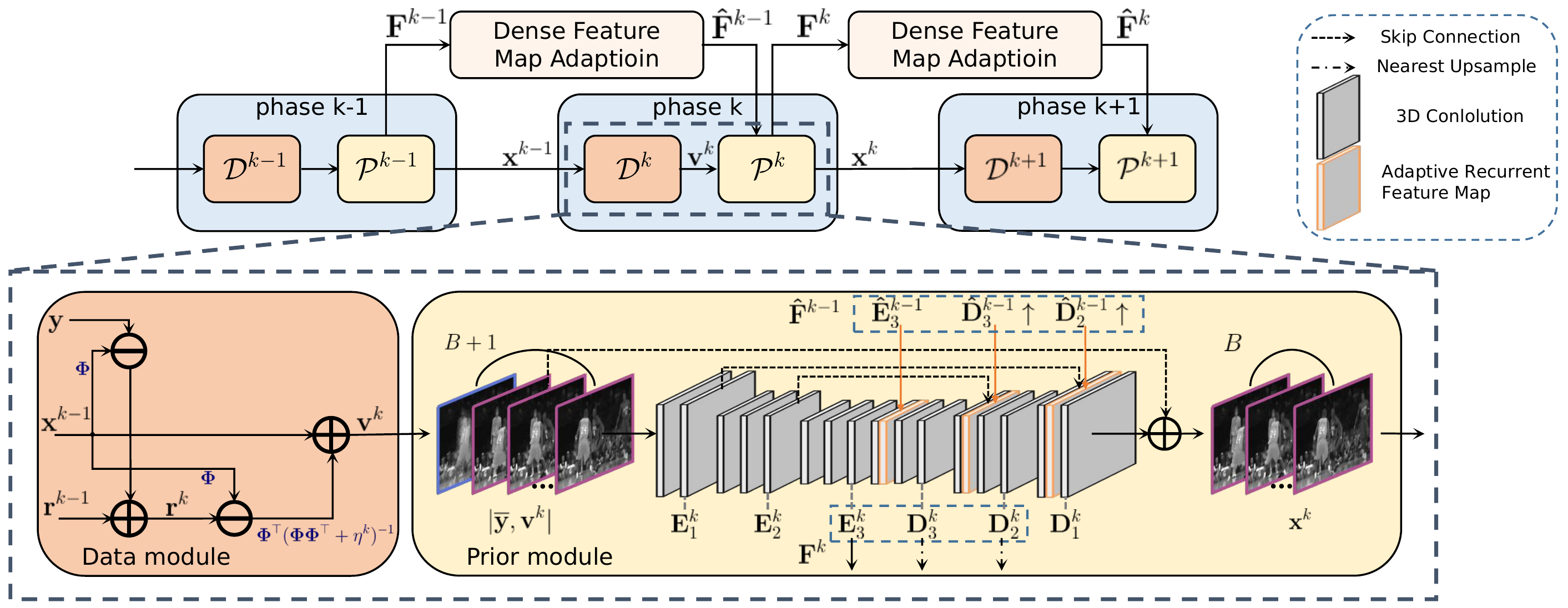}
% }
\caption{
Illustration of each phase in our proposed network. Data module solves Eqs.~{(\ref{9})(\ref{10})} and prior module solves Eq.~{(\ref{12})}. $\mathbf{F}^{k-1}$ represents dense feature map from the $(k-1)$-th phase and $\hat{\mathbf{F}}^{k-1}$ denotes its adaptive counterpart. The module of dense feature map adaption makes information selectively transmitted through phases. $\mathbf{E}_3^k$, $\mathbf{D}_3^k\uparrow$, and $\mathbf{D}_2^k\uparrow$ constitute the dense feature map $\mathbf{F}^k$.
}
\label{detail}
\end{figure*}

\subsection{Optimization-based Unrolled Network}
The SCI recovery problem can be solved by the following optimization problem:
\begin{equation}
    \mathbf{x} = \arg \min_\mathbf{x} \frac{1}{2}||\mathbf{y}-\boldsymbol{\Phi} \mathbf{x}||_2^2 + \lambda \Psi(\mathbf{x}), \label{4}
\end{equation}
where $\Psi(\mathbf{x})$ denotes the prior regularization with $\lambda$ being the regularization parameter.

Following the framework of HQS, by introducing an auxiliary parameter $\mathbf{v}$, Eq.~{(\ref{4})} can be converted to a constraint optimization problem:
\begin{equation}
    (\mathbf{x}, \mathbf{v}) = \arg \min_{\mathbf{x}, \mathbf{v}}\frac{1}{2}||\mathbf{y}-\boldsymbol{\Phi} \mathbf{v}||_2^2 + \lambda \Psi (\mathbf{x}), ~s.t.~ \mathbf{x}=\mathbf{v}. \label{5}
\end{equation}
The first term can be served as the data term, and the second term is the prior term. In order to obtain an unrolling inference, Eq.~{(\ref{5})} can be divided into the following sub-problems and solved iteratively:
\begin{equation}
    \mathbf{v}^{k} = \arg \min_\mathbf{v} \frac{1}{2}||\mathbf{y}-\boldsymbol{\Phi} \mathbf{v}||_2^2 + \frac{\eta}{2}||\mathbf{v}-\mathbf{x}^{k-1}||_2^2, \label{6}
\end{equation}
\begin{equation}
    \mathbf{x}^{k} = \arg \min_\mathbf{x} \frac{\eta}{2}||\mathbf{v}^{k} - \mathbf{x}||_2^2 + \lambda \Psi(\mathbf{x}). \label{7}
\end{equation}
Here, $k$ denotes the HQS iteration index, and $\eta$ is another regularizer.
% The $\mathbf{x}$ sub-problem is a special case of proximal mapping prox$_{\lambda/\eta, \phi (\mathbf{v}^k)}$ when $\phi(\mathbf{x}) = ||\Psi \mathbf{x}||_1$.

\begin{figure}[!t]
\centering 
% \subfigure[]{
\includegraphics[width=0.9\linewidth]{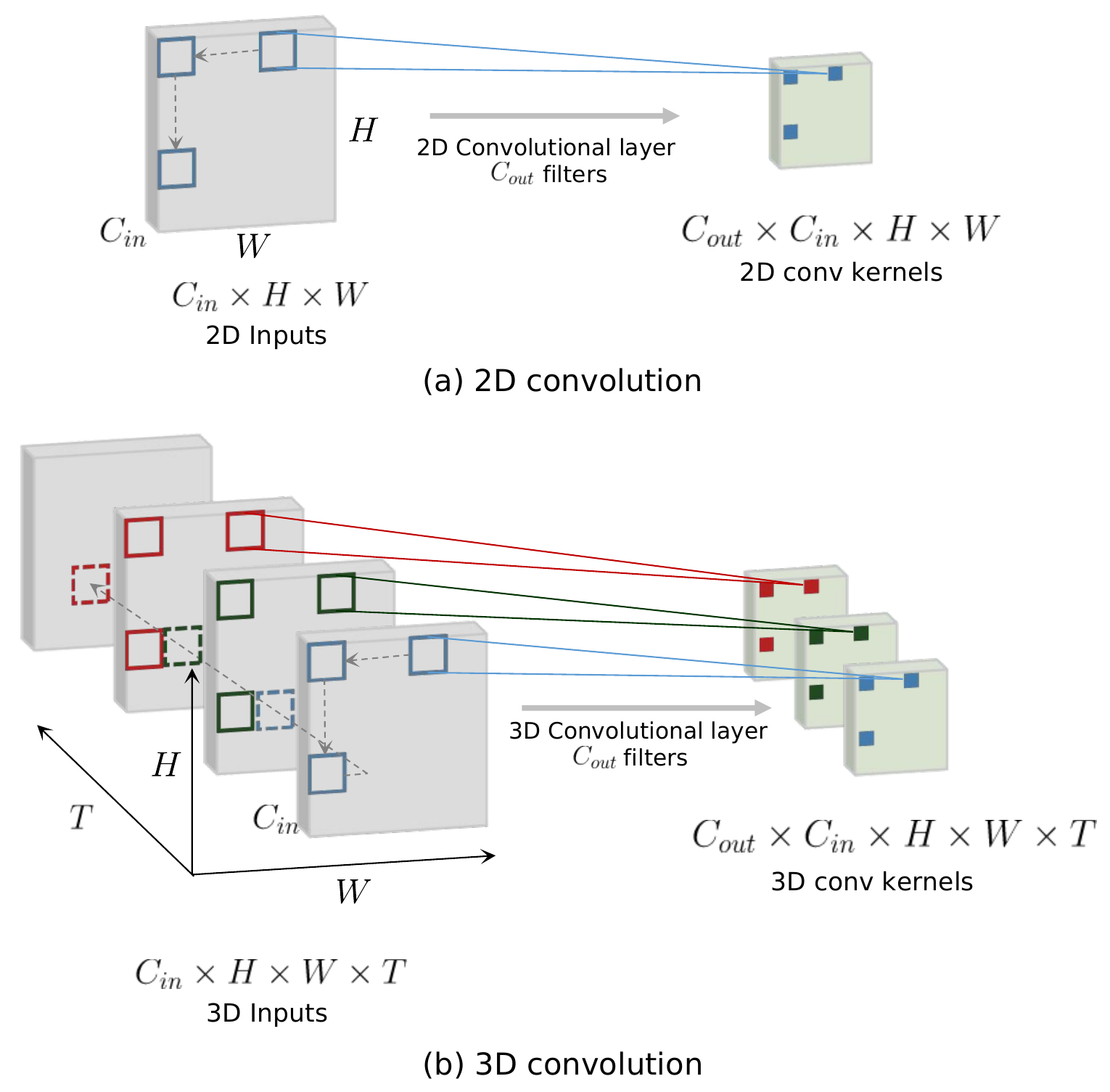}
% \subfigure[]{
% \includegraphics[width=1\textwidth]{detail_cropped_compressed.pdf}
% }
\caption{Illustration of comparison between 2D and 3D convolution. For 3D convolution, the input feature map has an additional dimension, which means filters of 3D convolution slide both along the spatial and temporal dimensions.}
\label{2Dvs3D}
\end{figure}

It can be observed that data term and prior term in Eq.~{(\ref{5})} are decoupled to sub-problems Eq.~{(\ref{6})} and Eq.~{(\ref{7})}. Intuitively, the aim of the data term and the prior term is to reconstruct frames $\mathbf{x}^k$ clearer and cleaner. Conventional optimization-based methods resolve the inverse problem by iteratively solving sub-problems according to optimization algorithms such as ADMM \cite{chan2016plug}, GAP \cite{liao2014generalized}, ISTA \cite{beck2009fast} and so on. However, it is prolonged for hundreds of iterations to obtain a decent reconstruction result. The basic idea of our proposed method is to unroll Eq.~{(\ref{6})} and Eq.~{(\ref{7})} into a deep network with a fixed number of phases, where each iteration corresponds to one phase. As shown in Fig.~{\ref{detail}}, $k$-th iteration of HQS is cast to $k$-th phase comprising data module $\mathcal{D}$ and prior module $\mathcal{P}$. Details will be discussed in the following:

\begin{itemize}
\item Data module $\mathcal{D}$: It corresponds to Eq.~(\ref{6}) and is used to generate the immediate reconstructed result $\mathbf{v}^k$. Given $\mathbf{x}^{k-1}$, the update of $\mathbf{v}$ in Eq.~{(\ref{6})} can be regarded as a euclidean projection of $\mathbf{v}$ on the linear manifold:
\begin{equation}
    \mathbf{v}^k = \mathbf{x}^{k-1} + \boldsymbol{\Phi}^\top(\boldsymbol{\Phi}\boldsymbol{\Phi}^\top + \eta^k)^{-1}(\mathbf{y}-\boldsymbol{\Phi}\mathbf{x}^{k-1}). \label{8}
\end{equation}
As shown in \cite{liao2014generalized}, the linear manifold can be adjusted adaptively, so Eq.~{(\ref{8})} can be modified as:
\begin{equation}
    \mathbf{r}^k = \mathbf{r}^{k-1} + (\mathbf{y}-\boldsymbol{\Phi} \mathbf{x}^{k-1}),~ \forall k\ge 1,
\label{9}
\vspace{-0.75cm}
\end{equation}
\begin{equation}
    \mathbf{v}^k = \mathbf{x}^{k-1} + \boldsymbol{\Phi}^\top(\boldsymbol{\Phi} \boldsymbol{\Phi}^\top + \eta^k)^{-1}(\mathbf{r}^{k}-\boldsymbol{\Phi} \mathbf{x}^{k-1}). 
\label{10}
\end{equation}
% \begin{equation}
% \begin{gathered}
%         \mathbf{r}^k = \mathbf{r}^{k-1} + (\mathbf{y}-\boldsymbol{\Phi} \mathbf{x}^{k-1}),~ \forall k\ge 1, \label{9}\\
%         \mathbf{v}^k = \mathbf{x}^{k-1} + \boldsymbol{\Phi}^\top(\boldsymbol{\Phi} \boldsymbol{\Phi}^\top + \eta^k)^{-1}(\mathbf{r}^{k}-\boldsymbol{\Phi} \mathbf{x}^{k-1}). \label{10}
% \end{gathered}
% \end{equation}
To increase the flexibility of network, $\eta^k$ for $k \in \{1,2,...,K\}$ is set as a separate learnable variable in each phase with initialization of 0.01. For simplification, the process of the data module can be described as:
\begin{equation}
        \mathbf{v}^k = \mathcal{D}^k(\mathbf{x}^{k-1}, \mathbf{r}^{k-1}, \eta^k, \mathbf{y}, \boldsymbol{\Phi}). \label{11}
\end{equation}
    
\item Prior module $\mathcal{P}$: The prior module aims to make $\mathbf{v}^k$ closer to the desired signal domain. It can be defined as:
\begin{equation}
        \mathbf{x}^k = \mathcal{P}^k(|\mathbf{v}^k, \overline{\mathbf{y}}|), \label{12}
\end{equation}
where $|\cdot|$ denotes the concatenation. For the prior module $\mathcal{P}^k$ in each phase, the input is the concatenate of $\mathbf{v}^k$ and $\overline{\mathbf{y}}$. As explained in Sec.~{\ref{Measurement Energy Normalization}}, the normalized measurement can provide visually pleasant motion information to guide reconstruction proceedings. Eq.~{(\ref{7})} is a special case of proximal mapping prox$_{\Psi, \lambda/\eta}$ associated with the non-linear transform $\Psi(\cdot)$. Traditionally the sparse transformation $\Psi(\cdot)$ is set by handcraft. Inspired by \cite{yuan2020plug}, we design a deep network to solve the problem of proximal mapping.
% we design a deep network aiming to seek the help from the strong ability of representation of CNN. 

Our network shares the similar architecture of U-Net as designed in \cite{tassano2020fastdvdnet}. However, we apply 3D convolution layers instead of the 2D ones. Fig.~{\ref{2Dvs3D}} gives an illustration of the differences between 2D and 3D convolution. By contrast with filters in 2D convolution, 3D ones sweep in additional dimension (\textit{i.e.}, temporal duration), making it more suitable to exploit spatial-temporal correlation, an intuitional idea that is proved by our later experiments.

As portrayed in Fig.~{\ref{detail}}, the architecture of the prior module $\mathcal{P}^k$ in the $k$-th phase is a modified U-Net consisting of an encoder and a decoder. Considering there are three scales in both encoder along with decoder and architecture in each scale is equivalent to a residual block, we use $j\in [1,2,3]$ to denote the scale where $j=1$ represents the largest scale, thereby the output of residual block in separate scale can be denoted as $\mathbf{E}_j^k$ for the encoder and $\mathbf{D}_j^k$ for the decoder. The procedure of the whole network can be expressed as  
\begin{equation}
\begin{gathered}
    \mathbf{E}_j^k = \text{enc}_{j}(\mathbf{E}_{j-1}^k), \\
    \mathbf{D}_j^k = \text{dec}_j(\overline{\mathbf{D}}_j),
\end{gathered}
\end{equation}
 where $\overline{\mathbf{D}}_j$ is the input for decoder at level $j$, details will be discussed in Sec.~{\ref{DFM}}. It should be noticed that $\mathbf{E}_0^k = |\mathbf{v}^k, \mathbf{\overline{y}}|$.

 Finally, the prior module learns a residual map:
%  which means that the input of the prior module, and the latter term indicates there is no skip connection between encoder and decoder in the smallest scale (bottom)
\begin{equation}
    \mathbf{x}^k = \mathbf{D}_1^k + \mathbf{v}^k. \label{15}
\end{equation}

\end{itemize}

\vspace{-0.2cm}
\subsection{Dense Feature Map Fusion}
\label{DFM}
First, we are going to discuss the bottleneck in existing DUNs. As shown in Fig.~{{\ref{SCI}}}, the input of reconstruction network $\mathbf{x}^0$ is assigned with $\boldsymbol{\Phi^\top \mathbf{y}}$. Then, each phase contains data module $\mathcal{D}$ and prior module $\mathcal{P}$ cast from Eqs.~{(\ref{9})(\ref{10})} and Eq.~{(\ref{12})}. 

In the data module, for some intermediate result $\mathbf{u}$ the notation of $\boldsymbol{\Phi} \mathbf{u}$ represents the operation of image-wise sampling while $\boldsymbol{\Phi}^\top \mathbf{u}$ denotes the operation of image-wise zero-filling. Given $\mathbf{x}^{k-1} \in \mathbb{R}^{B\times H\times W}$, after corresponding image-wise sample \& zero-fill operation according to Eqs.~{(\ref{9})(\ref{10})}, we get $\mathbf{v}^k$, then in the prior module, $\mathbf{x}^k$ is often obtained by solving a problem of the proximal mapping via a designed denoising network.

It is obvious that the input and output of each phase contain limited channels, that is, each frame of $\mathbf{v}^k$ and $\mathbf{x}^k$ is simply containing minor channels (\textit{e.g.}, 1 for gray images and 3 for RGB images). There exist information loss during transmissions between phases, which greatly hinders the performance of the entire network. To break the bottleneck, we come up with a strategy of dense feature map (DFM) fusion. Our basic idea can be exhibited as follows
\begin{equation}
\begin{gathered}
    \mathbf{D}_3^k = \text{dec}_3(|\mathbf{E}_3^k, \mathbf{E}_3^{k-1}|), \\
    \mathbf{D}_2^k = \text{dec}_2(|\mathbf{D}_3^k\uparrow + \mathbf{E}_2^k, \mathbf{D}_3^{k-1}\uparrow|), \\
    \mathbf{D}_1^k = \text{dec}_1(|\mathbf{D}_2^k\uparrow + \mathbf{E}_1^k, \mathbf{D}_2^{k-1}\uparrow|) \label{16}
\end{gathered}
\end{equation}
where $\uparrow$ denotes the nearest up-sample and feature maps from $(k-1)$-th phase are named as dense feature map
\begin{equation}
    \mathbf{F}^{k-1} = [\mathbf{E}_3^{k-1}, \mathbf{D}_3^{k-1}\uparrow, \mathbf{D}_2^{k-1}\uparrow]. \label{17}
\end{equation}
We fuse dense feature map from the beginning of the decoder at each scale because, particularly for such architecture as encoder-decoder, information passed through the encoder has been compressed and necessary to provide extra information for the decoder.

% We select to fuse inter-phase information from the beginning of decoder at each level since for such networks constructed by encoder-decoder, information passing through encoder has been compressed. So it is necessary to provide extra information for decoder. Specifically, at $k$-th phase denoting $\widetilde{\mathbf{F} }_i^k$ as the input for decoder at level $i$ and it can be obtained by

% Coming from the output of ResBlock in corresponding $i$-th level of $(k-1)$-th phase, dense feature map $F^{k-1}_i$ is concatenated with the up-asmpled feature map $f^k_i$ generated from previous ResBlock of $i$-th level in $k$-th phase as the input of decoder $\widetilde{F}^k_i$ in level $i$:
% \begin{equation}
%     \widetilde{\mathbf{F}}^k_i = |\mathbf{F}^{k-1}_i, \mathbf{f}^k_i|, \label{15}
% \end{equation}
% where $\mathbf{F}_i^{k-1}$ is dense feature map from $(k-1)$-th phase and $\mathbf{f}^k_i$ is up-sampled feature map generated from previous level.

\begin{figure}[!t]
\centering 
% \subfigure[]{
\includegraphics[width=0.9\linewidth]{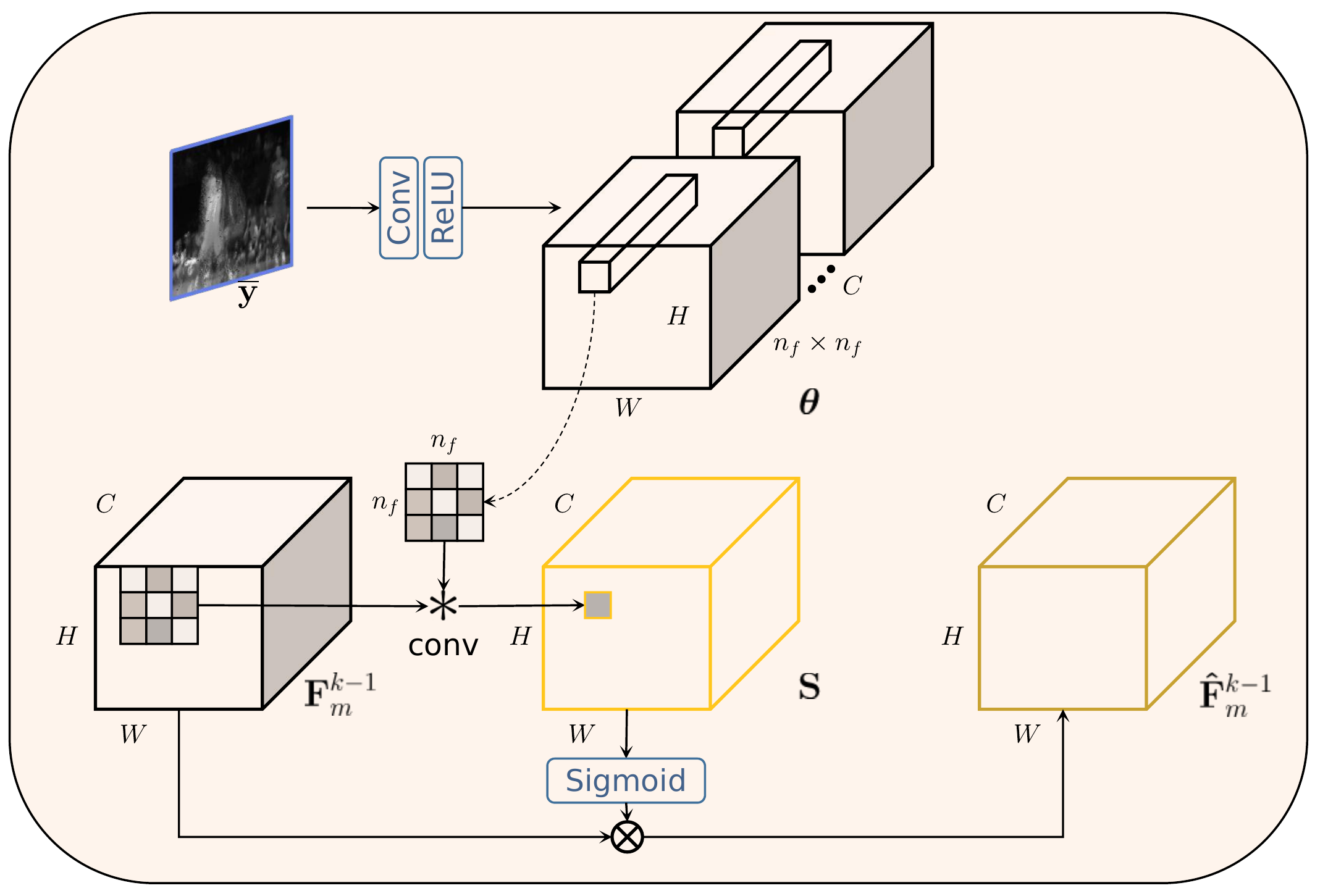}
% \subfigure[]{
% \includegraphics[width=1\textwidth]{detail_cropped_compressed.pdf}
% }
\caption{Visualization of our dense feature map adaption (DFMA) module. Similarity matrix $\mathbf{S}$ is derived by convolving a plain $m$-th DFM $\mathbf{F}_m^{k-1}$ where kernels $\theta$ is generated through a convolution from normalized measurement $\mathbf{\overline{y}}$. Adaptive DFM is obtained by multiplying $\mathbf{S}$ and $\mathbf{F}_m^{k-1}$.}
\label{DFMA}
\end{figure}

\begin{table*}[t]
\caption{Comparison with other algorithms. The average results of PSNR in dB (left entry in each cell) and SSIM (right entry in each cell) and running time per measurement in seconds. Best results are in \textcolor{red}{red} and the second-best ones are in \textcolor{blue}{blue}.}
\label{tab: table 1}
\resizebox{\textwidth}{60pt}{

\begin{tabular}{lcccccccc}
	\hline
		Dataset & Kobe & Traffic & Runner & Drop & Aerial & Crash & Average & Running time \\
	\hline
		GAP-TV \cite{yuan2016generalized}  & 26.45 0.845  & 20.90 0.715 & 28.48 0.899 & 33.81 0.963 & 25.03 0.828 & 24.82 0.838 & 26.58 0.848 & 4.2 \\
		E2E-CNN \cite{qiao2020deep} & 27.79 0.807 & 24.62 0.840 & 34.12 0.947 & 36.56 0.949 & 27.18 0.869 & 26.43 0.882 & 29.45 0.882 & 0.0312 \\
		DeSCI \cite{liu2018rank} & 33.25 0.952 & 28.72 0.925 & 38.76 0.969 & \textcolor{blue}{43.22 0.993} & 25.33 0.860 & 27.04 0.909 & 32.72 0.935 & 6180 \\
		PnP-FFDNet \cite{yuan2020plug} & 30.47 0.926 & 24.08 0.833 & 32.88 0.938 & 40.87 0.988 & 24.02 0.814 & 24.32 0.836 & 29.44 0.889 & 3.0 \\
		BIRNAT \cite{yuanbirnat} & 32.71 0.950 & 29.33 0.942 & 38.70 0.976 & 42.28 0.992 & 28.99 0.927 & 27.84 0.927 & 33.31 0.951 & 0.16 \\
		Tensor-ADMM \cite{ma2019deep} & 30.50 0.890 & NA & NA & NA & 25.42 0.780 &  25.27 0.860 & NA & 2.1 \\
		Tensor-FISTA \cite{han2020tensor} & 31.41 0.920 & NA & NA & NA & 26.46 0.890 & 27.46 0.880 & NA & 1.7 \\
		GAP-Unet-S12 \cite{meng2020gap} & 32.09 0.944 &  28.19 0.929 & 38.12 0.975 & 42.02 0.992 & 28.88 0.914 & 27.83 0.931 & 32.86 0.947 & 0.0072 \\
		
		MetaSCI \cite{wang2021metasci} & 30.12 0.907 & 26.95 0.888 & 37.02 0.967 & 40.61 0.985 & 28.31 0.904 & 27.33 0.906 & 31.72 0.926 & 0.025 \\
		RevSCI \cite{cheng2021memoryefficient} & \textcolor{blue}{33.72 0.957} & \textcolor{blue}{30.02 0.949} & \textcolor{blue}{39.40 0.977} & 42.93 0.992 & \textcolor{blue}{29.35 0.924} & \textcolor{blue}{28.12 0.937} & \textcolor{blue}{33.92 0.956} & 0.19 \\
% 	\bottomrule
% 	    spatial & 31.74 0.9375 & 27.02 0.9125 & 36.12 0.9638 & 42.25 0.9914 & 28.35 0.9043 & 27.73 0.9169 & 32.20 0.9377 & 0.2304 \\
% 	    temporal & 33.63 0.9571 & 29.72 0.9494 & 37.76 0.9720 & 43.58 0.9935 & 29.68 0.9292 & 28.57 0.9382 & 33.83 0.9566 & 0.3209 \\
% 	    spatial + temporal & 33.75 0.9568 & \textcolor{red}{29.98 0.9519} & 38.05 0.9726 & \textcolor{red}{43.84 0.9938} & 29.73 0.9306 & 28.58 0.9373 & 33.99 0.9572 & 0.5661 \\

% % 	\bottomrule
	   % \textbf{Ours} &\textcolor{red}{34.59 0.966} & \textcolor{red}{31.32 0.963} &\textcolor{red}{39.50 0.979} &\textcolor{red}{44.67 0.9945} &\textcolor{red}{30.25 0.940} & \textcolor{red}{29.32 0.955} & \textcolor{red}{34.94 0.966}  & 1.1799 \\
	    \textbf{Ours} &\textcolor{red}{35.00 0.969} & \textcolor{red}{31.76 0.966} &\textcolor{red}{40.03 0.980} &\textcolor{red}{44.96 0.995} &\textcolor{red}{30.46 0.943} & \textcolor{red}{29.33 0.956} & \textcolor{red}{35.26 0.968}  & 1.35 \\
	    
	\hline
\end{tabular}}
\end{table*}

\subsection{Dense Feature Map Adaption Module}
% For low-level vision tasks, there have been vigorous developing for long-range information fusion while dense neural network (RNN) is one of the most widely applicated one. The previous RNN-based SCI recovery method BIRNAT \cite{yuanbirnat} employs bidirectional RNN to reconstruct frames in a sequential manner. For such RNN-based methods, past information is transmitted to current time step via hidden state by concatenating with certain feature map. 
As described in Sec.~{\ref{DFM}}, notwithstanding that our proposed DFM fusion has decrease information loss to a great extent, we realize that applying the same integration manner to all DFMs is not an optimal way. Intuitively, different channels of DFM are able to contribute differently to final fusion. Inspired by \cite{isobe2020video}, we design a dense feature map adaption (DFMA) module to adaptively fuse inter-phase information.

Our assumption is that each unit in DFM $\mathbf{F}^{k-1}$ which exhibits a similar appearance with the normalized measurement $\mathbf{\overline{y}}$ should be highlighted while others are supposed to be suppressed. In this way, only useful information in DFM can be transmitted between phases. As depicted in Fig.~{\ref{DFMA}}, our basic idea is calculating a similarity map $\mathbf{S}$. As mentioned in \cite{JiaBTG16}, one can generate spatially variant filters and samples on each location to compute the similarity between DFM and normalized measurement. More concretely, spatially variant filters $\boldsymbol{\theta} \in \mathcal{R}^{H\times W\times C\times n_f\times n_f}$ are generated from normalized measurement $\overline{\mathbf{y}}$, where $H$, $W$ and $C$ are the height, length and channels of corresponding scale $j$. $n_f$ is the kernel size, which is being set as $3$ in our experiment. The similarity $\mathbf{S}(p,q,c)$ between $\mathbf{F}_{m}^{k-1}$ and $\overline{\mathbf{y}}$ at location $(p,q)$ in $c$-th channel can be computed by
\begin{equation}
\begin{gathered}
    \mathbf{S}(p,q,c) = \\
    \sum_{u=-z}^{z} \sum_{v=-z}^{z} \boldsymbol{\theta}(p,q,c,u,v) \times \mathbf{F}_m^{k-1}(p+u, q+v, c),
\end{gathered}
\end{equation}
where $z = \lfloor n_f/2 \rfloor$ and $\mathbf{F}_{m}^{k-1}$ signifies the $m$-th ($m=1,2,3$) entry of DFM $\mathbf{F}^{k-1}$. Then adapted dense feature map $\hat{\mathbf{F}}_m^{k-1}$ is obtained by
\begin{equation}
    \hat{\mathbf{F}}_m^{k-1} = \sigma(\mathbf{S}) \otimes \mathbf{F}_{m}^{k-1},
\end{equation}
where $\sigma$ denotes the sigmoid function to convert its value into the range $[0,1]$. Then Eq.~{(\ref{17})} can be modified to
\begin{equation}
    \hat{\mathbf{F}}^{k-1} = [\hat{\mathbf{E}}_3^{k-1}, \hat{\mathbf{D}}_3^{k-1}\uparrow, \hat{\mathbf{D}}_2^{k-1}\uparrow].
\end{equation}

% Correspondingly, Eq.~{(\ref{16})} can be modified with
% \begin{equation}
%     \widetilde{\mathbf{F}}^k_i = |\hat{\mathbf{F}}^{k-1}_i, \mathbf{f}^k_i|.
% \end{equation}

\begin{algorithm}[t] %固定位置 !h
	\caption{Dense Deep Unfolding Network with 3D-CNN Prior for SCI}%算法标题
    \hspace*{0.02in} {\bf Input: Sensing matrix $\boldsymbol{\Phi}$, observed measurement $\mathbf{y}$}
    
    \hspace*{0.02in} {\bf Output: Reconstructed frames $\mathbf{x}^K$}
    
	\begin{algorithmic}[1]%一行一个标行号
	    \STATE \text{Initialize $\mathbf{x}^0 = \boldsymbol{\Phi}^\top\mathbf{y}$, $\mathbf{r}^0 = \bf{0}$}
	    \FOR{$k=1,...,K$}
	    \STATE \text{update $\mathbf{r}^k$ by Eq.~(\ref{9})}
	    \STATE \text{update $\mathbf{v}^k$ by Eq.~(\ref{10})}
	    \STATE \text{update $\mathbf{x}^k$ by Eq.~(\ref{15})}
	    \ENDFOR 
% 		\STATE $i=p$
% 		\FOR{$j=p$ to $r$}
% 		\IF{$A[j]<=0$}
% 		\STATE $swap(A[i],A[j])$
% 		\STATE $i=i+1$
% 		\ENDIF
% 		\ENDFOR
	\end{algorithmic}
\end{algorithm}

\subsection{Training}
The learnable parameters are denoted as $\Theta$, including $\eta^k$ in the data module and parameters of the deep network in the prior module $\mathcal{P}^k$. Thus $\Theta = \{\eta^k, \mathcal{P}^k\}_{k=1}^K$, where $K$ is total number of phases ($K=10$ in our proposed network).

Given the training pairs $\{(\mathbf{y}_i, \mathbf{x}_i)\}_{i=1}^{N_k}$, our network takes the measurement $\mathbf{y}_i$ as input and generates reconstructed frames $\mathbf{x}_i^K$. The mean square error (MSE) is chosen as our loss function, expressed as
\begin{equation}
    \mathcal{L} = \frac{1}{N_kN_s}\sum_{i=1}^{N_k} ||\mathbf{x}_i^{K} - \mathbf{x}_i||_2^2,
\end{equation}
where $N_k$, $N_s$ are the total number of training blocks and size of each block $\mathbf{x}_i$.

The architecture is implemented in PyTorch \cite{NEURIPS2019_9015} with 4 NVIDIA Tesla V100 GPUs. To optimize the parameters, we adopt the Adam optimizer \cite{kingma2014adam} with a mini-batch size of 4 and the number of total epoch is 200. The initial learning rate is set to be $1.28\times 10^{-4}$, but we find it is too large to be converged. To overcome early optimization difficulties, we use the strategy of warmup \cite{he2016deep} with the first five epochs, \textit{i.e.}, using lower learning rates at the start of training. After the first five epochs, the learning rate decays with a factor of 0.9 every ten epochs.

\section{Experiments}
\subsection{Datasets}
We choose \textbf{DAVIS} \cite{Pont-Tuset_arXiv_2017} which has 90 scenes with 480p resolution as our training dataset. 25600 video clips are sampled with the block size of $128\times 128$ and the temporal size of 8. Data is augmented with random cropping and rotation. The synthetic data with size of $256\times 256\times 8$ follows the same setup in \cite{liu2018rank}, including \textbf{Kobe, Traffic, Runner, Drop, Crash,} and \textbf{Aerial} \cite{ma2019deep}. The real data includes \textbf{Water Balloon} and \textbf{Dominoes} with size of $512\times 512\times 10$ \cite{qiao2020deep}. Both PSNR and SSIM \cite{wang2004image} are selected as the evaluation metrics.

\subsection{Results on Simulation Data}
We compare our proposed method with existing algorithms including model-based methods (GAP-TV \cite{yuan2016generalized} and DeSCI \cite{liu2018rank}), learning-based methods (E2E-CNN \cite{qiao2020deep}, BIRNAT \cite{yuanbirnat}, MetaSCI \cite{wang2021metasci} and RevSCI \cite{cheng2021memoryefficient}) and DUN-based methods (PnP-FFDNet \cite{yuan2020plug}, Tensor-ADMM \cite{ma2019deep}, Tensor-FISTA \cite{han2020tensor} and GAP-Unet-S12 \cite{meng2020gap}) on synthetic data. It should be noted that for Tensor-ADMM and Tensor-FISTA only 3 results out of 6 scenes are provided because they only trained models on 3 scenes. Quantitative comparison is shown in Table~\ref{tab: table 1}. 

\begin{figure}[!t]
\centering 
% \subfigure[]{
% \includegraphics[width=1\textwidth]{SCI_cropped.pdf}}
% \subfigure[]{
\includegraphics[width=0.95\linewidth]{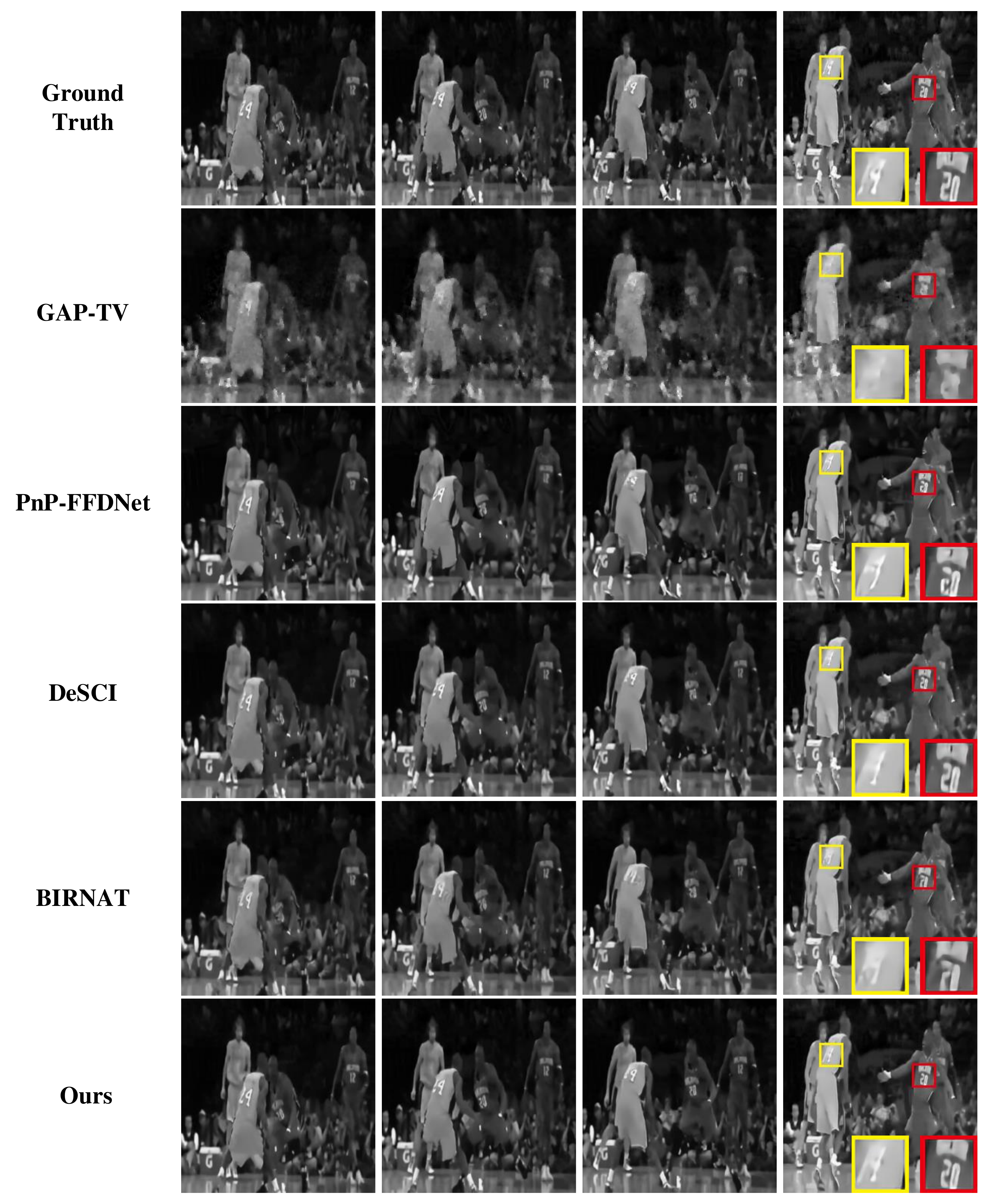}
% }
\caption{Selected multiple reconstruction frames of simulated data \textbf{Kobe}.}
\label{kobe}
\end{figure}

\begin{figure}[!t]
\centering 
% \subfigure[]{
% \includegraphics[width=1\textwidth]{SCI_cropped.pdf}}
% \subfigure[]{
\includegraphics[width=1\linewidth]{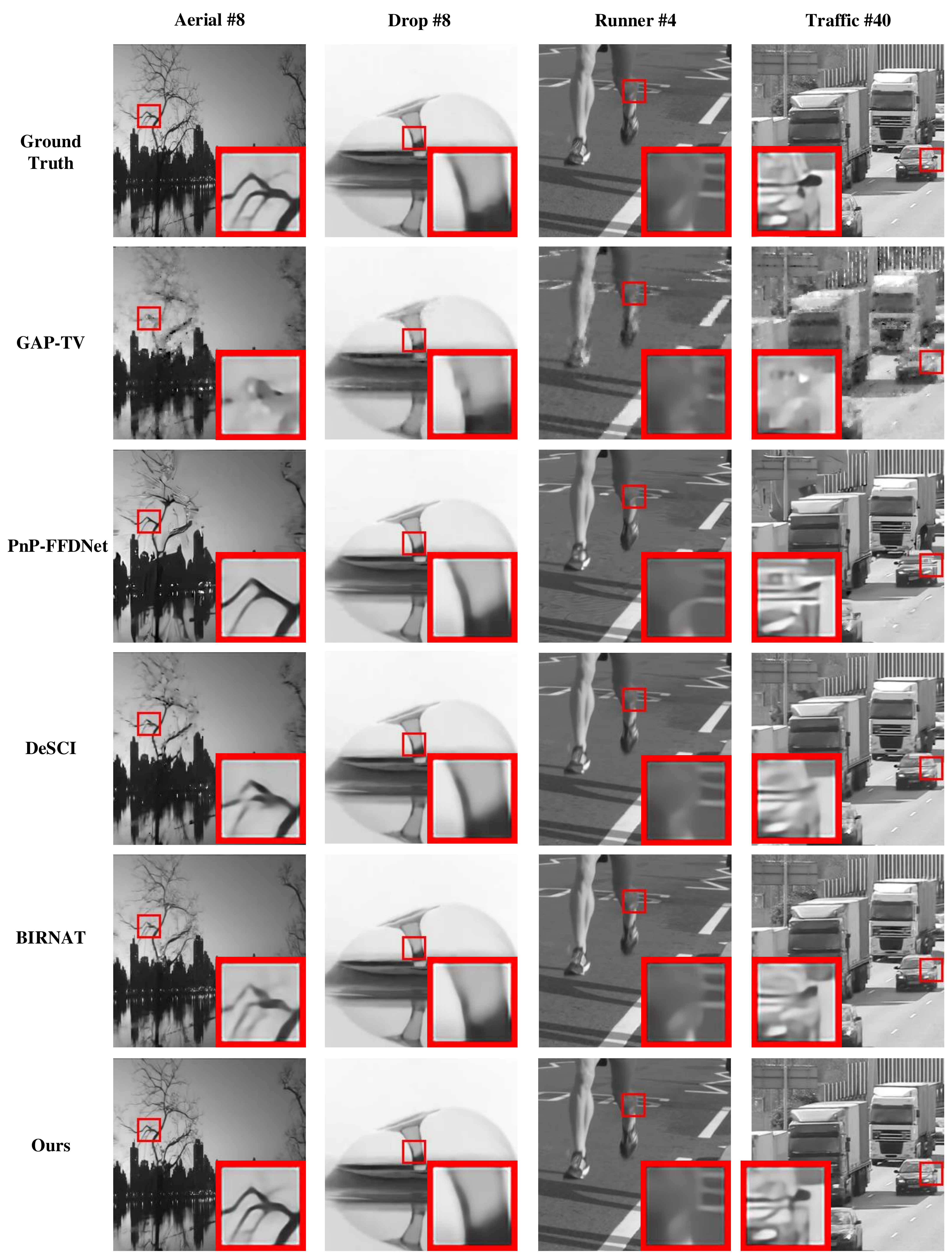}
% }
\caption{Selected reconstruction frames of simulated data \textbf{Aerial, Drop, Runner and Traffic}.}
\label{simulation_data}
\end{figure}

It is obvious that our proposed network has surpassed all the existing algorithms on six benchmarks by a large margin. Especially our performance exceeds the existing best algorithm RevSCI by 1.34dB on average. Fig.~{\ref{kobe}} and Fig.~{\ref{simulation_data}}  plots some selected reconstructed results on simulation data. Note that the model-based method GAP-TV has worse reconstruction results. Built on the plug-and-play framework, PnP-FFDNet obtains decent results but still lacks details, especially in texture-rich areas. The results of DeSCI are a little bit over smooth with the cost of high computation load. Compared with BIRNAT, in zooming areas, our method obtains sharper edges and more abundant details, demonstrating the effectiveness of utilization of 3D-CNN and the success of 
declining information loss.

% For brevity, we simply exhibit visual results on Kobe and Traffic. As shown in Fig.~{\ref{kobe}} and Fig.~{\ref{traffic}}, combined with quantitive comparison results, we can provide more details and our reconstructed frames are visually clearer (as displayed in zooming boxes).

% \subsection{Results on Large-scale Data}
\subsection{Results on Real Data}
\begin{figure}[!t]
\centering 
% \subfigure[]{
% \includegraphics[width=1\textwidth]{SCI_cropped.pdf}}
% \subfigure[]{
\includegraphics[width=1\linewidth]{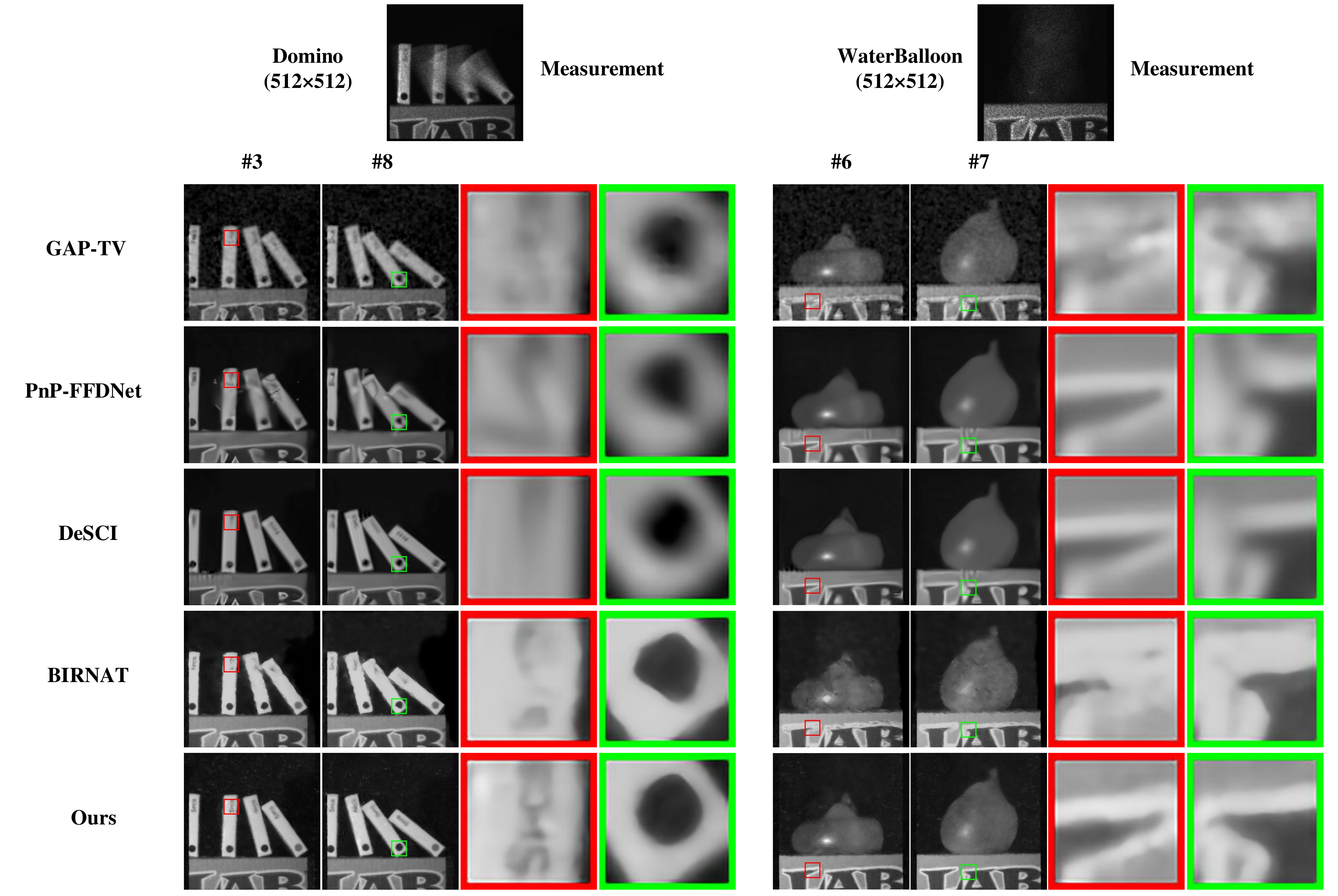}
% }
\caption{Selected reconstruction frames of real data.}
\label{real_data}
\end{figure}

We apply our network to real data $\mathbf{Water Balloon}$ and $\mathbf{Dominoes}$ \cite{qiao2020deep} where masks are controlled by a digital micromirror device. It is more challenging to reconstruct the real measurements because of noise. As illustrated in Fig.~{\ref{real_data}}, our method can generate more apparent contours with fewer artifacts and ghosting. In general, our method achieves the best performance with competing inference speed. When facing a more challenging situation, our method's advantages are more pronounced, which indicates that our algorithm is more applicable under the actual scenario.

\subsection{Ablation Study}

% \begin{table}[t]
% \caption{Architecture of denoising block.}
% \label{tab: table 2}
% \begin{threeparttable}
% \begin{tabular}{c|cccc}
% 	\hline
% 		Layer No. & type      & kernel   & stride   & channel  \\
% 	\hline
% 		1 & input     &    -     & -        & c\tnote{*}   \\
% 		  & Conv + Relu & $3\times3$ & $1\times1$ & 30c \\
% 		  & Conv + Relu & $3\times3$ & $1\times1$ & 32  \\
% 		\hline
% 		2 & Conv$\downarrow$ + Relu & $3\times3$ & $2\times2$ & 64 \\
% 		  & Conv + Relu & $3\times3$ & $1\times1$ & 64 \\
% 		  & Conv + Relu & $3\times3$ & $1\times1$ & 64 \\
% 		\hline
% 		3 & Conv$\downarrow$ + Relu & $3\times3$ & $2\times2$ & 128 \\
% 		  & Conv + Relu & $3\times3$ & $1\times1$ & 128 \\
% 		  & Conv + Relu & $3\times3$ & $1\times1$ & 128 \\
% 		\hline
% 		4 & Conv + Relu & $3\times3$ & $1\times1$ & 128 \\
% 		  & Conv + Relu & $3\times3$ & $1\times1$ & 128 \\
% 		  & Conv      & $3\times3$ & $1\times1$ & 256 \\
% 		  & PixelShuffle &  -    &  -       & 64  \\
% 		\hline
% 		5 & Conv + Relu & $3\times3$ & $1\times1$ & 64 \\
% 		  & Conv + Relu & $3\times3$ & $1\times1$ & 64 \\
% 		  & Conv      & $3\times3$ & $1\times1$ & 128 \\
% 		  & PixelShuffle &  -    &  -       & 32  \\
% 		\hline
% 		6 & Conv + Relu & $3\times3$ & $1\times1$ & 32 \\
% 		  & Conv      & $3\times3$ & $1\times1$ & 1 \\
% 		\hline
% \end{tabular}
% \begin{tablenotes}    
%         \footnotesize
%         \item[*] For denoising block in spatial step $c=1$ while in temporal step $c=3$.          
%       \end{tablenotes}  
% \end{threeparttable}
% \end{table}

\begin{table}[t]
\caption{Ablation study on convolution: 3D-CNN versus 2D-CNN.}
\label{tab: table 2}
\begin{center}
\begin{tabular}{ccc}
    \hline
    Sequence & 2DCN & 3DCN \\
    \hline
    Kobe &32.07 0.940 & 34.09 0.962 \\
    Traffic &28.26 0.931 & 30.72 0.958 \\
    Runner &36.43 0.967 & 38.81 0.976 \\
    Drop &42.38 0.992 & 44.20 0.994\\
    Aerial &29.18 0.921 & 30.04 0.936\\
    Crash &28.29 0.934 & 29.06 0.950\\
    \hline
    Average & 32.77 0.948 & 34.49 0.963 \\
    \hline
\end{tabular}
\end{center}
\end{table}

\begin{table}[t]
\caption{Ablation study on various versions in terms of the average results of PSNR in dB.}
\label{tab: table 3}
\centering
\begin{tabular}{cccccc}
	\hline
		3DCN & $DFM_1$ & $DFM_2$ & $DFM_3$ & DFMA & PSNR  \\
	\hline
	    \checkmark & & & & & 34.49  \\
	    \checkmark & \checkmark & & & & 34.54  \\
	    \checkmark & \checkmark & & &\checkmark & 34.81  \\
	    \checkmark & \checkmark & \checkmark & & & 34.80  \\
	    \checkmark & \checkmark & \checkmark & &\checkmark & 35.03  \\
	    \checkmark & \checkmark & \checkmark & \checkmark & & 34.94 \\
	    \checkmark & \checkmark & \checkmark & \checkmark & \checkmark & 35.26\\
	    
	\hline
\end{tabular}
\end{table}

% \begin{figure*}[!t]
% \centering 
% % \subfigure[]{
% % \includegraphics[width=1\textwidth]{SCI_cropped.pdf}}
% % \subfigure[]{
% \includegraphics[width=0.65\linewidth]{ICCV_visual_comparison_cropped.pdf}
% % }
% \caption{Selected reconstruction frames of simulation data.}
% \label{simulation_data}
% \end{figure*}

\subsubsection{Effect of 3D-CNN over 2D-CNN}
\label{3D-CNN}
To validate the effectiveness of the application of 3D-CNN, we trained a plain 3D modified UNet (`3DCN') \cite{tassano2020fastdvdnet} and corresponding 2D version (`2DCN') without integration of the DFM fusion and DFMA module. As shown in Table~\ref{tab: table 2}, 3DCN can achieve an increment of 1.72dB on PSNR, which shows not just sliding in the dimension of height and width, sliding in temporal dimension is necessary for extracting spatial-temporal correlation.

\subsubsection{Effect of DFM and DFMA}
To verify the proposed DFM fusion and DFMA, we integrate diverse branches based on 3DCN in Sec.~{\ref{3D-CNN}} with three branches in $\mathbf{F}^k$ denoted as `DFM$_1$', `DFM$_2$' and `DFM$_3$' respectively. As manifested in Table~{\ref{tab: table 3}}, combining the second branch improves the performance most (0.26dB) contrast with the first (0.05dB) and third branch (0.14dB). The whole branches bring an improvement of 0.45dB, proving that merging dense feature map from adjacent phases can reduce the information loss to a great extent. Incorporating DFMA into each branch brings additional enhancement, and the whole branch of DFMA leads to a boost of 0.32dB on PSNR.

\subsubsection{Effect of Phase Number $K$}
\begin{figure}[!t]
\centering 
% \subfigure[]{
% \includegraphics[width=1\textwidth]{SCI_cropped.pdf}}
% \subfigure[]{
\includegraphics[width=0.9\linewidth]{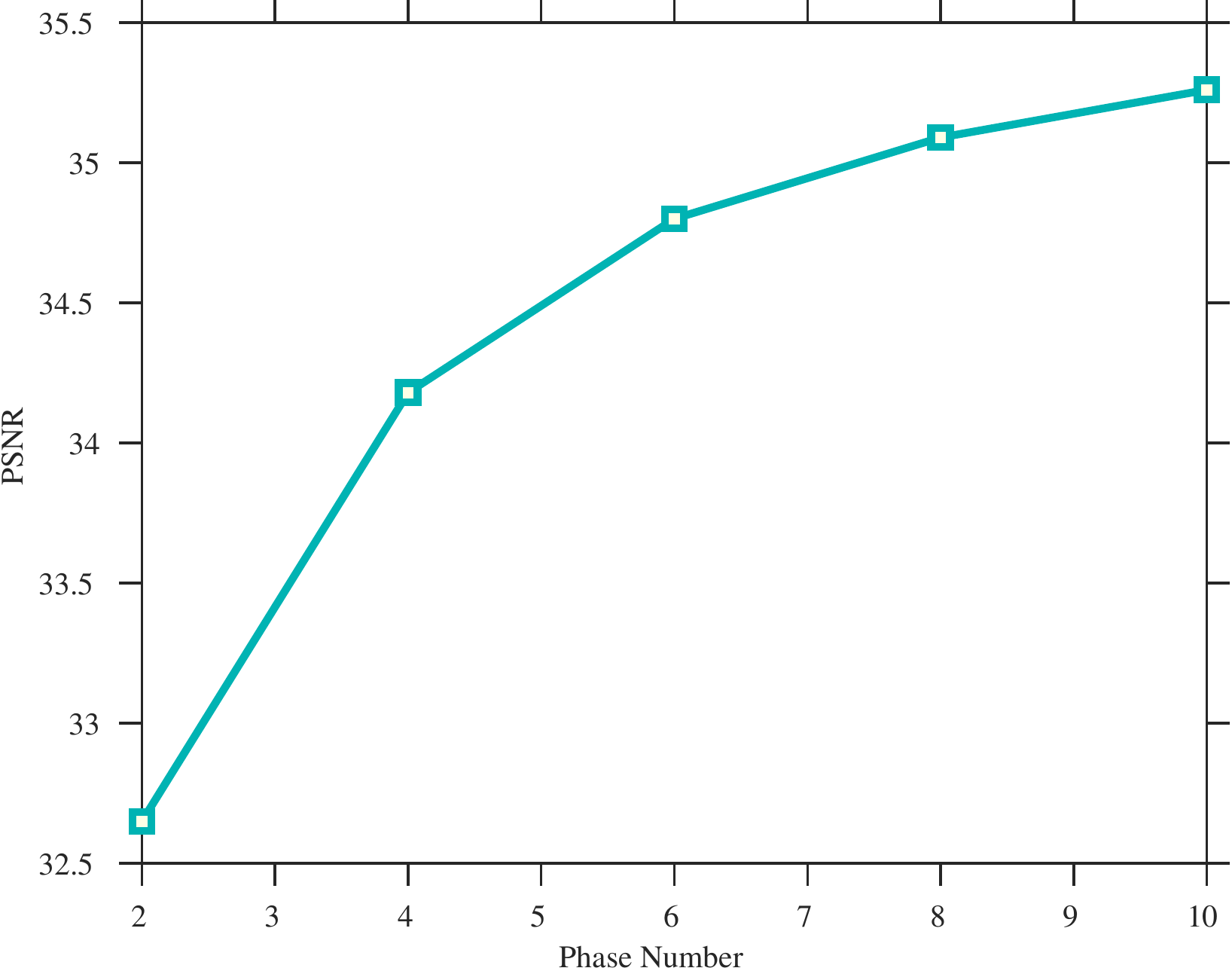}
% }
\caption{PSNR comparison on reconstruction accuracy with various number of phase.}
\label{phase-PSNR}
\end{figure}

To ensure the phase number $K$ being set as an appropriate value, we retrain the network with $K$ being designated to $2,4,6,8$ and $10$. Fig.~{\ref{phase-PSNR}} plots the performance with various phase numbers. Note that the performance with $K=4$ is 0.26dB higher than RevSCI. Although reconstruction accuracy will increase with more phases, considering the memory cost and inference time, we assign $K$ to be 10.

\begin{figure}[!t]
\centering 
% \subfigure[]{
% \includegraphics[width=1\textwidth]{SCI_cropped.pdf}}
% \subfigure[]{
\includegraphics[width=1\linewidth]{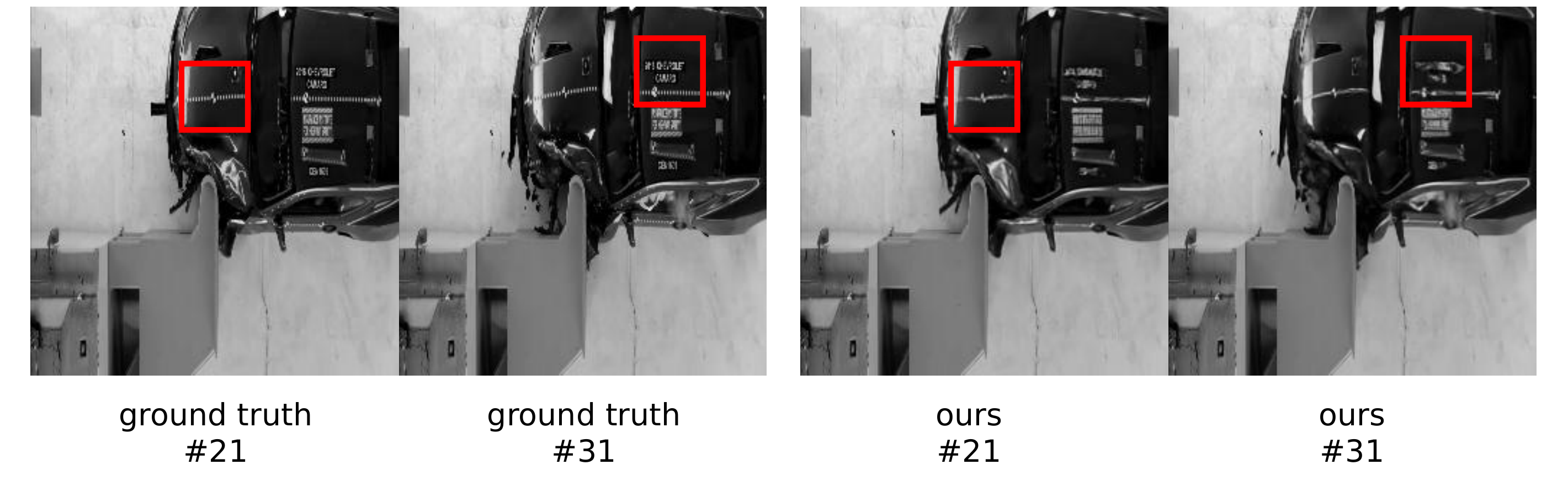}
% }
\caption{A case of unsatisfactory reconstruction results due to large motion.}
\label{failure}
\end{figure}

\section{Conclusions and Future Work}
Inspired by the half quadratic splitting (HQS) algorithm, we put forward a novel dense deep unfolding network with 3D-CNN prior for Snapshot compressive imaging. Merging the merits of both model-based methods and learning-based methods, our method has strong interpretability and high-quality reconstructed results. To enhance the ability to exploit spatial-temporal correlation, we assemble a deep network with 3D-CNN prior. To reduce the information loss, we propose a strategy of dense feature map (DFM) fusion, and we also design a dense feature map adaption (DFMA) module to make information optionally transmitting between phases.

Despite the fact that our method has obtained the best performance so far, when it comes to large motion within frames, our method can not attain decent results, as shown in Fig.~{\ref{failure}}. How to properly process large motion is thorny trouble for existing SCI recovery methods. In our future work, we might insert motion estimation techniques such as optical flow to solve this trouble. In addition, we will consider the application in other computational imaging problems, not just video SCI.

\newpage
{\small
\bibliographystyle{ieee_fullname}
\bibliography{egbib}

\begin{thebibliography}{10}\itemsep=-1pt

\bibitem{beck2009fast}
Amir Beck and Marc Teboulle.
\newblock A fast iterative shrinkage-thresholding algorithm for linear inverse
  problems.
\newblock {\em SIAM Journal on Imaging Sciences}, 2(1):183--202, 2009.

\bibitem{chan2016plug}
Stanley~H Chan, Xiran Wang, and Omar~A Elgendy.
\newblock Plug-and-play admm for image restoration: Fixed-point convergence and
  applications.
\newblock {\em IEEE Transactions on Computational Imaging}, 3(1):84--98, 2016.

\bibitem{chang2019free}
Ya-Liang Chang, Zhe~Yu Liu, Kuan-Ying Lee, and Winston Hsu.
\newblock Free-form video inpainting with 3d gated convolution and temporal
  patchgan.
\newblock In {\em Proceedings of the IEEE/CVF International Conference on
  Computer Vision}, pages 9066--9075, 2019.

\bibitem{cheng2021memoryefficient}
Ziheng Cheng, Bo Chen, Guanliang Liu, Hao Zhang, Ruiying Lu, Zhengjue Wang, and
  Xin Yuan.
\newblock Memory-efficient network for large-scale video compressive sensing.
\newblock In {\em Proceedings of the IEEE/CVF Conference on Computer Vision and
  Pattern Recognition (CVPR)}, pages 16246--16255, June 2021.

\bibitem{yuanbirnat}
Ziheng Cheng, Ruiying Lu, Zhengjue Wang, Hao Zhang, Bo Chen, Ziyi Meng, and Xin
  Yuan.
\newblock Birnat: Bidirectional recurrent neural networks with adversarial
  training for video snapshot compressive imaging.
\newblock In {\em Proceedings of the IEEE conference on European Conference on
  Computer Vision (ECCV)}, 2020.

\bibitem{cciccek20163d}
{\"O}zg{\"u}n {\c{C}}i{\c{c}}ek, Ahmed Abdulkadir, Soeren~S Lienkamp, Thomas
  Brox, and Olaf Ronneberger.
\newblock 3d u-net: learning dense volumetric segmentation from sparse
  annotation.
\newblock In {\em Proceedings of the International conference on Medical Image
  Computing and Computer-assisted Intervention}, pages 424--432. Springer,
  2016.

\bibitem{duarte2008single}
Marco~F Duarte, Mark~A Davenport, Dharmpal Takhar, Jason~N Laska, Ting Sun,
  Kevin~F Kelly, and Richard~G Baraniuk.
\newblock Single-pixel imaging via compressive sampling.
\newblock {\em IEEE Signal Processing Magazine}, 25(2):83--91, 2008.

\bibitem{geman1995nonlinear}
Donald Geman and Chengda Yang.
\newblock Nonlinear image recovery with half-quadratic regularization.
\newblock {\em IEEE transactions on Image Processing}, 4(7):932--946, 1995.

\bibitem{han2020tensor}
Xiaochen Han, Bo Wu, Zheng Shou, Xiao-Yang Liu, Yimeng Zhang, and Linghe Kong.
\newblock Tensor fista-net for real-time snapshot compressive imaging.
\newblock In {\em Proceedings of the AAAI Conference on Artificial
  Intelligence}, volume~34, pages 10933--10940, 2020.

\bibitem{he2016deep}
Kaiming He, Xiangyu Zhang, Shaoqing Ren, and Jian Sun.
\newblock Deep residual learning for image recognition.
\newblock In {\em Proceedings of the IEEE Conference on Computer Vision and
  Pattern Recognition}, pages 770--778, 2016.

\bibitem{iliadis2018deep}
Michael Iliadis, Leonidas Spinoulas, and Aggelos~K Katsaggelos.
\newblock Deep fully-connected networks for video compressive sensing.
\newblock {\em Digital Signal Processing}, 72:9--18, 2018.

\bibitem{isobe2020video}
Takashi Isobe, Xu Jia, Shuhang Gu, Songjiang Li, Shengjin Wang, and Qi Tian.
\newblock Video super-resolution with recurrent structure-detail network.
\newblock In {\em Proceedings of the European Conference on Computer Vision},
  pages 645--660. Springer, 2020.

\bibitem{jalali2019snapshot}
Shirin Jalali and Xin Yuan.
\newblock Snapshot compressed sensing: Performance bounds and algorithms.
\newblock {\em IEEE Transactions on Information Theory}, 65(12):8005--8024,
  2019.

\bibitem{ji20123d}
Shuiwang Ji, Wei Xu, Ming Yang, and Kai Yu.
\newblock 3d convolutional neural networks for human action recognition.
\newblock {\em IEEE Transactions on Pattern Analysis and Machine Intelligence},
  35(1):221--231, 2012.

\bibitem{JiaBTG16}
Xu Jia, Bert De~Brabandere, Tinne Tuytelaars, and Luc~V Gool.
\newblock Dynamic filter networks.
\newblock {\em Proceedings of the Advances in Neural Information Processing
  Systems}, 29:667--675, 2016.

\bibitem{kim20183dsrnet}
Soo~Ye Kim, Jeongyeon Lim, Taeyoung Na, and Munchurl Kim.
\newblock 3dsrnet: Video super-resolution using 3d convolutional neural
  networks.
\newblock {\em arXiv preprint arXiv:1812.09079}, 2018.

\bibitem{kingma2014adam}
Diederik~P Kingma and Jimmy Ba.
\newblock Adam: A method for stochastic optimization.
\newblock {\em arXiv preprint arXiv:1412.6980}, 2014.

\bibitem{li2020end}
Yuqi Li, Miao Qi, Rahul Gulve, Mian Wei, Roman Genov, Kiriakos~N Kutulakos, and
  Wolfgang Heidrich.
\newblock End-to-end video compressive sensing using anderson-accelerated
  unrolled networks.
\newblock In {\em Proceedings of the 2020 IEEE International Conference on
  Computational Photography (ICCP)}, pages 1--12. IEEE, 2020.

\bibitem{liao2014generalized}
Xuejun Liao, Hui Li, and Lawrence Carin.
\newblock Generalized alternating projection for weighted-2,1 minimization with
  applications to model-based compressive sensing.
\newblock {\em SIAM Journal on Imaging Sciences}, 7(2):797--823, 2014.

\bibitem{liu2018rank}
Yang Liu, Xin Yuan, Jinli Suo, David~J Brady, and Qionghai Dai.
\newblock Rank minimization for snapshot compressive imaging.
\newblock {\em IEEE Transactions on Pattern Analysis and Machine Intelligence},
  41(12):2990--3006, 2018.

\bibitem{llull2013coded}
Patrick Llull, Xuejun Liao, Xin Yuan, Jianbo Yang, David Kittle, Lawrence
  Carin, Guillermo Sapiro, and David~J Brady.
\newblock Coded aperture compressive temporal imaging.
\newblock {\em Optics Express}, 21(9):10526--10545, 2013.

\bibitem{luo2020video}
Jianping Luo, Shaofei Huang, and Yuan Yuan.
\newblock Video super-resolution using multi-scale pyramid 3d convolutional
  networks.
\newblock In {\em Proceedings of the 28th ACM International Conference on
  Multimedia}, pages 1882--1890, 2020.

\bibitem{ma2019deep}
Jiawei Ma, Xiao-Yang Liu, Zheng Shou, and Xin Yuan.
\newblock Deep tensor admm-net for snapshot compressive imaging.
\newblock In {\em Proceedings of the IEEE/CVF International Conference on
  Computer Vision}, pages 10223--10232, 2019.

\bibitem{maggioni2012video}
Matteo Maggioni, Giacomo Boracchi, Alessandro Foi, and Karen Egiazarian.
\newblock Video denoising, deblocking, and enhancement through separable 4-d
  nonlocal spatiotemporal transforms.
\newblock {\em IEEE Transactions on Image Processing}, 21(9):3952--3966, 2012.

\bibitem{meng2020gap}
Ziyi Meng, Shirin Jalali, and Xin Yuan.
\newblock Gap-net for snapshot compressive imaging.
\newblock {\em arXiv preprint arXiv:2012.08364}, 2020.

\bibitem{NEURIPS2019_9015}
Adam Paszke, Sam Gross, Francisco Massa, Adam Lerer, James Bradbury, Gregory
  Chanan, Trevor Killeen, Zeming Lin, Natalia Gimelshein, Luca Antiga, et~al.
\newblock Pytorch: An imperative style, high-performance deep learning library.
\newblock {\em Proceedings of the Advances in Neural Information Processing
  Systems}, 32:8026--8037, 2019.

\bibitem{Pont-Tuset_arXiv_2017}
Jordi Pont-Tuset, Federico Perazzi, Sergi Caelles, Pablo Arbel\'aez, Alexander
  Sorkine-Hornung, and Luc {Van Gool}.
\newblock The 2017 davis challenge on video object segmentation.
\newblock {\em arXiv:1704.00675}, 2017.

\bibitem{qiao2020deep}
Mu Qiao, Ziyi Meng, Jiawei Ma, and Xin Yuan.
\newblock Deep learning for video compressive sensing.
\newblock {\em APL Photonics}, 5(3):030801, 2020.

\bibitem{reddy2011p2c2}
Dikpal Reddy, Ashok Veeraraghavan, and Rama Chellappa.
\newblock P2c2: Programmable pixel compressive camera for high speed imaging.
\newblock In {\em Proceedings of the IEEE Conference on Computer Vision and
  Pattern Recognition}, pages 329--336. IEEE, 2011.

\bibitem{tassano2020fastdvdnet}
Matias Tassano, Julie Delon, and Thomas Veit.
\newblock Fastdvdnet: Towards real-time deep video denoising without flow
  estimation.
\newblock In {\em Proceedings of the IEEE Conference on Computer Vision and
  Pattern Recognition}, pages 1354--1363, 2020.

\bibitem{teivas2017video}
Iikka~Tapio Teivas.
\newblock Video event classification using 3d convolutional neural networks.
\newblock Master's thesis, 2017.

\bibitem{tran2015learning}
Du Tran, Lubomir Bourdev, Rob Fergus, Lorenzo Torresani, and Manohar Paluri.
\newblock Learning spatiotemporal features with 3d convolutional networks.
\newblock In {\em Proceedings of the IEEE International Conference on Computer
  Vision}, pages 4489--4497, 2015.

\bibitem{wagadarikar2009video}
Ashwin~A Wagadarikar, Nikos~P Pitsianis, Xiaobai Sun, and David~J Brady.
\newblock Video rate spectral imaging using a coded aperture snapshot spectral
  imager.
\newblock {\em Optics Express}, 17(8):6368--6388, 2009.

\bibitem{wang2019video}
Chuan Wang, Haibin Huang, Xiaoguang Han, and Jue Wang.
\newblock Video inpainting by jointly learning temporal structure and spatial
  details.
\newblock In {\em Proceedings of the AAAI Conference on Artificial
  Intelligence}, volume~33, pages 5232--5239, 2019.

\bibitem{wang2004image}
Zhou Wang, Alan~C Bovik, Hamid~R Sheikh, and Eero~P Simoncelli.
\newblock Image quality assessment: from error visibility to structural
  similarity.
\newblock {\em IEEE Transactions on Image Processing}, 13(4):600--612, 2004.

\bibitem{wang2021metasci}
Zhengjue Wang, Hao Zhang, Ziheng Cheng, Bo Chen, and Xin Yuan.
\newblock Metasci: Scalable and adaptive reconstruction for video compressive
  sensing.
\newblock In {\em Proceedings of the IEEE/CVF Conference on Computer Vision and
  Pattern Recognition}, pages 2083--2092, 2021.

\bibitem{wu2021spatial}
Zhuoyuan Wu, Zhenyu Zhang, Jiechong Song, and Jian Zhang.
\newblock Spatial-temporal synergic prior driven unfolding network for snapshot
  compressive imaging.
\newblock In {\em Proceedings of IEEE International Conference on Multimedia
  and Expo (ICME)}, 2021.

\bibitem{yang2014video}
Jianbo Yang, Xin Yuan, Xuejun Liao, Patrick Llull, David~J Brady, Guillermo
  Sapiro, and Lawrence Carin.
\newblock Video compressive sensing using gaussian mixture models.
\newblock {\em IEEE Transactions on Image Processing}, 23(11):4863--4878, 2014.

\bibitem{yoshida2018joint}
Michitaka Yoshida, Akihiko Torii, Masatoshi Okutomi, Kenta Endo, Yukinobu
  Sugiyama, Rin-ichiro Taniguchi, and Hajime Nagahara.
\newblock Joint optimization for compressive video sensing and reconstruction
  under hardware constraints.
\newblock In {\em Proceedings of the European Conference on Computer Vision
  (ECCV)}, pages 634--649, 2018.

\bibitem{you2021}
Di You, Jian Zhang, Jingfen Xie, Bin Chen, and Siwe Ma.
\newblock {COAST: COntrollable Arbitrary-Sampling NeTwork for Compressive
  Sensing}.
\newblock {\em IEEE Transactions on Image Processing}, 30:6066--6080, 2021.

\bibitem{yuan2016generalized}
Xin Yuan.
\newblock Generalized alternating projection based total variation minimization
  for compressive sensing.
\newblock In {\em Proceedings of IEEE International Conference on Image
  Processing (ICIP)}, pages 2539--2543. IEEE, 2016.

\bibitem{yuan2020plug}
Xin Yuan, Yang Liu, Jinli Suo, and Qionghai Dai.
\newblock Plug-and-play algorithms for large-scale snapshot compressive
  imaging.
\newblock In {\em Proceedings of the IEEE Conference on Computer Vision and
  Pattern Recognition}, pages 1447--1457, 2020.

\bibitem{zhang2018ista}
Jian Zhang and Bernard Ghanem.
\newblock {ISTA-Net:} interpretable optimization-inspired deep network for
  image compressive sensing.
\newblock In {\em Proceedings of the IEEE Conference on Computer Vision and
  Pattern Recognition}, pages 1828--1837, 2018.

\bibitem{zhangopine}
Jian Zhang, Chen Zhao, and Wen Gao.
\newblock Optimization-inspired compact deep compressive sensing.
\newblock {\em IEEE Journal of Selected Topics in Signal Processing},
  14(4):765--774, 2020.

\bibitem{zhao2014image}
Chen Zhao, Siwei Ma, and Wen Gao.
\newblock Image compressive-sensing recovery using structured laplacian
  sparsity in dct domain and multi-hypothesis prediction.
\newblock In {\em Proceedings of IEEE International Conference on Multimedia
  and Expo (ICME)}, 2014.

\bibitem{zhao2016video}
Chen Zhao, Siwei Ma, Jian Zhang, Ruiqin Xiong, and Wen Gao.
\newblock Video compressive sensing reconstruction via reweighted residual
  sparsity.
\newblock {\em IEEE Transactions on Circuits and Systems for Video Technology},
  27(6):1182--1195, 2016.

\bibitem{zhao2018cream}
Chen Zhao, Jian Zhang, Ronggang Wang, and Wen Gao.
\newblock {CREAM: CNN-REgularized} admm framework for compressive-sensed image
  reconstruction.
\newblock {\em IEEE Access}, 6:76838--76853, 2018.

\end{thebibliography}
}

\end{document}